\documentclass[preprint,groupedaddress,floatfix,nofootinbib]{revtex4}
\usepackage{euscript}
\usepackage{dcolumn}% Align table columns on decimal point
\usepackage{graphicx}
\usepackage{epsfig}
\usepackage{hyperref}
\usepackage{graphicx}% Include figure files
\usepackage{dcolumn}% Align table columns on decimal point
\usepackage{longtable}
\usepackage{times}
\usepackage{amsmath,amssymb,bm}
\usepackage[english]{babel}
\usepackage[latin1]{inputenc}
\usepackage{times}
\usepackage[T1]{fontenc}
\usepackage{color}
\usepackage{fancybox}
\usepackage{sidecap}
\usepackage{graphicx}
\usepackage{epsfig}
\usepackage{psfrag}
\usepackage{bbm}
\usepackage[english]{babel}
\usepackage[latin1]{inputenc}
\usepackage{times}
\usepackage[T1]{fontenc}
\usepackage{color}
\usepackage{fancybox}
\usepackage{sidecap}
\usepackage{psfrag}
\usepackage{bbm}
\usepackage{wasysym}
\usepackage{slashed}
\usepackage{tikz}
\usepackage{euscript}
\usepackage{booktabs}
\usepackage{algorithm}
\usepackage{algpseudocode}
\usepackage{float}

\begin{document}

\title{Improving stabilizer approximation with quantum strategy}

\author{Fen Zuo\footnote{Email: \textsf{zuofen@miqroera.com}}}
\affiliation{Hefei MiQro Era Digital Technology Co. Ltd., Hefei, China}

\begin{abstract}
We introduce a quantum strategy from nonlocal games to improve the stabilizer approximation we proposed previously. The resulting approach turns out to be a qubit-by-qubit gauging procedure for standard stabilizers, which could involve discrete or continuous gauge parameters. We take examples from many-body physics and quantum chemistry to show such a procedure leads to an improvement of the performance.
\end{abstract}
 \maketitle

\tableofcontents

\section{A brief introduction}

The famous toric code~\cite{Kitaev97} presents a beautiful connection between a stabilizer code~\cite{Gottesman-1996} and a quantum many-body problem: we could use the stabilizers to engineer a Hamiltonian, such that its ground states constitute exactly the logical subspace, or the stabilized subspace. Reversing the logic, we would like to get the ground states of some Hamiltonian with stabilizers. For realistic Hamiltonians, we wouldn't be so luck to obtain the exact ground states anymore. So we settle for approximate ground states. For Hamiltonians expressed as sums of Pauli terms, a practical approach would be to select an appropriate commuting subset of Pauli terms as the stabilizers, such that the subspace they stabilize would be an approximate ground state. This is the so-called stabilizer approximation that we proposed in \cite{SA-I} and developed in \cite{SA-II,SA-III}, based on previous studies~\cite{Shang-2021,CAFQA}. See \cite{Cheng-2022,Sun-2024,Brown-2024} for related investigations.

While stabilizer approximation performs rather well in most cases, it could go bad when two non-commuting terms possess coefficients of close magnitude. In such a situation choosing either operator as the stabilizer would not be good enough, as discussed in~\cite{Cheng-2022}. This in fact should have been anticipated. Stabilizer formalism alone is not adequate to handle the Hamiltonian problem, in the same spirit that Clifford circuits could not provide universal quantum computation~\cite{Gottesman-1997,Gottesman-1998}. To do better, we need to introduce more ``quantumness'' into it.

The development of nonlocal games, especially the Clause-Horne-Shimony-Holt~(CHSH) game~\cite{CHSH}, gives us a clue as how to achieve a quantum advantage. So we start with the CHSH game, and try to borrow the proper quantum strategy from it to improve our approximation.

\section{The CHSH game}
We follow the presentation of the CHSH game in Thomas Vidick's lecture notes~\cite{Vidick-2018}. Two players in the game are given binary queries $i$ and $j$ respectively, and give also binary answers $a$ and $b$. They win if the answers satisfy
\begin{equation}
a_i\oplus b_j=i\wedge j,
\end{equation}
where $\oplus$ is logical XOR, $\wedge$ is logical AND. Assuming the queries are uniformly chosen, we could easily calculate the classical bias and winning probability. To do this, we linearly transform the binary values of $a$ and $b$ into parity/spin values $\{+1,-1\}$, but still use the same labels to denote them. Then, we have for the bias
\begin{equation}
\beta_c=\max \frac{1}{4}(a_0b_0+a_0b_1+a_1b_0-a_1b_1)=\frac{1}{2},
\end{equation}
and the winning probability
\begin{equation}
\omega_c=\frac{1}{2}+\frac{1}{2}\beta_c=\frac{3}{4}.
\end{equation}
So classically we have at most a chance of $\frac{3}{4}$ to win the game.

Quantumly, we could do strictly better. We promote $\{a_0,a_1\}$ and $\{b_0,b_1\}$ into
\begin{equation}
A_0=Z,~A_1=X;\quad B_0=H=\frac{1}{\sqrt{2}}(X+Z),~B_1=\tilde H\equiv \frac{1}{\sqrt{2}}(Z-X).\label{eq.HH}
\end{equation}
All of them have eigenvalues $\{+1,-1\}$. Now we let the two players share an EPR pair:
\begin{equation}
|\Phi\rangle=\frac{|00\rangle+|11\rangle}{\sqrt{2}},
\end{equation}
and take the values of the respective measurements $\{A_i, B_j\}$ as their answers. Therefore, the quantum bias reads
\begin{equation}
\beta_q=\frac{1}{4}\langle \Phi|A_0\otimes B_0+A_0\otimes B_1+A_1\otimes B_0-A_1\otimes B_1|\Phi\rangle=\frac{\sqrt{2}}{2}.
\end{equation}
The winning probability of the quantum strategy is then
\begin{equation}
 \omega_q=\frac{1}{2}+\frac{1}{2}\beta_q=\frac{1}{2}+\frac{\sqrt{2}}{4}\approx 0.853,
 \end{equation}
which is indeed greater the classical value. Moreover, it is proved to be the best one.

\section{CHSH vs. stabilizer}

The CHSH game has a nice interpretation in terms of the stabilizer formalism, as shown in~\cite{Ji-2016}. In particular, it is shown that one could encode all the information of the game into a single Hamiltonian, and seek its ground state for the answer. The recipe is as follows. For each question instance $ij$, one includes a Pauli term into the Hamiltonian, with a $X$ factor for the ``0'' question, and a $Z$ factor for the ``1'' question. The expected answer, $i\wedge j$, is encoded into the coefficient of the Pauli term. Explicitly, one transforms it into a spin value, and sets the coefficient as its opposite. Doing so, we get the CHSH Hamiltonian~\cite{Ji-2016}:
%We slightly modify the notations and presentation in~\cite{Ji-2016} to show the significance for the ground state problem.
%Say we have a Hamiltonian:
\begin{equation}
H_0=-X\otimes X-X\otimes Z-Z\otimes X+Z\otimes Z.
\end{equation}
Now we try to approximate its ground state with stabilizer states. We could choose a commuting subset of terms, say $X\otimes X,-Z\otimes Z$, as stabilizers. The corresponding stabilizer state is
\begin{equation}
|\Psi\rangle=\frac{|01\rangle+|10\rangle}{\sqrt{2}}.
\end{equation}
And the corresponding energy, or expectation value of the Hamiltonian, is $-2$. The previous quantum strategy inspires us to make the following transformation:
\begin{eqnarray}
X'&=&R^\dagger_y(\frac{\pi}{4})~X~R_y(\frac{\pi}{4}), \\
Z'&=&R^\dagger_y(\frac{\pi}{4})~Z~R_y(\frac{\pi}{4}),
\end{eqnarray}
with
\begin{equation}
R_y(\theta)=\cos(\frac{\theta}{2})I-\mathrm{i}\sin(\frac{\theta}{2})Y.
\end{equation}
Essentially, we are taking the rotation around the $\hat y$ axis by $\pi/4$. Explicitly, this gives
\begin{equation}
X'=\frac{X+Z}{\sqrt{2}}=H,\quad Z'=\frac{Z-X}{\sqrt{2}}=\tilde H,
\end{equation}
as in eq.(\ref{eq.HH}). Then we could rewrite the Hamiltonian as
\begin{equation}
H_0=-\sqrt{2}(X\otimes X'-Z\otimes Z').
\end{equation}
Thus if we choose $\{X\otimes X',-Z\otimes Z'\}$ as our new stabilizers, we get a lower energy $-2\sqrt{2}$. In fact, this is the lowest eigenvalue of $H_0$. The corresponding new stabilizer state is simply~\cite{Ji-2016}
\begin{equation}
|\Psi'\rangle = I\otimes R_y^\dagger(\frac{\pi}{4})|\Psi\rangle.
\end{equation}
It is not difficult to check that $X\otimes X'$ and $-Z\otimes Z'$ indeed stabilize $|\Psi'\rangle$.
 Therefore, the quantum strategy used in the CHSH game helps us improve the stabilizer approximation for the ground state energy. The essential point behind the strategy is that, for different qubits we could use completely different sets of Pauli operators to define the stabilizers. In other words, we could gauge the Pauli operators for each qubit individually, with the single-qubit rotations. I am not sure whether ``gauge'' is the proper word here to describe the procedure exactly, but I prefer to use it anyhow.

\section{Gauging stabilizers}

\subsection{Individual gauging}

Now we apply the above procedure to some realistic problems, to see whether it is of help more generally. The first example is the Ising model with both longitudinal and transverse magnetic fields~\cite{Cheng-2022}:
\begin{equation}
H_{\rm{Ising}}=\sum_{ij} J Z_i\otimes Z_j+\sum_i (g_x X_i+g_z Z_i).
\end{equation}
When $g_x\sim g_z\gg J$, we would have difficulty to select our stabilizers properly. Taking an extremal situation, $g_x=g_z=1$, $J=0$, we have
\begin{equation}
H'_{\rm{Ising}}=\sum_i(X_i+Z_i).
\end{equation}
At each site, we may choose either $-Z_i$ or $-X_i$ as our stabilizer. This gives an energy $-N$, with $N$ the number of sites. Now we gauge the Pauli operators uniformly as before, and get
\begin{equation}
H'_{\rm{Ising}}=\sqrt{2}\sum_i X'_i.
\end{equation}
Choosing $-X_i'$ as the new stabilizer at each site, we get a lower energy $-\sqrt{2}N$, which is actually the lowest eigenvalue of $H'_{\rm{Ising}}$. Similar discussion for the generalized toric code model with external magnetic fields is done in~\cite{Sun-2024}.

\subsection{Sequential gauging}

The above example looks a little bit trivial. Now we give a slightly involved example. Consider the electronic Hamiltonian of the hydrogen molecule. We transform it into the qubit form with a proper transformation, say the parity transformation~\cite{Parity}. In the bounding region, the Hamiltonian looks like this
\begin{equation}
H_{H_2}=2I_1\otimes Z_2-2Z_1\otimes I_2.
\end{equation}
We could easily select the stabilizers as $-I_1\otimes Z_2, Z_1\otimes I_2$, and get the corresponding state
\begin{equation}
|\Psi_{HF}\rangle =|01\rangle.
\end{equation}
This is a product/Hartree-Fock~(HF) state, and the corresponding energy $-4$ is exact. In the asymptotic region, the correlation/resonating term dominates, and the Hamiltonian looks like:
\begin{equation}
H_{H_2}'=2X_1\otimes X_2
\end{equation}
We would get a single stabilizer $-X_1\otimes X_2$, which gives a degenerate state space with the exact energy $-2$. In the region in between, the Hamiltonian would be certain combination of the above two, say
\begin{equation}
H_{H_2}''= I_1\otimes Z_2-Z_1\otimes I_2+2X_1\otimes X_2.
\end{equation}
Choosing either the HF stabilizers $-I_1\otimes Z_2, Z_1\otimes I_2$ or the correlation stabilizer $-X_1\otimes X_2$ gives an energy $-2$. But this is not so satisfiable, because either choice ignores completely the contributions from the alternative side.

To properly include contributions from both sets of stabilizers, we run the following gauging procedure. First, we gauge $X_1$ and $Z_1$ as before, and discard $Z_1'$, say, by choosing $X_1'$ as the first stabilizer. We are left with the reduced Hamiltonian:
\begin{equation}
H_{H_2}''\sim -\frac{1}{\sqrt{2}}I_2+Z_2+\sqrt{2}X_2.
\end{equation}
Now we gauge $X_2$ and $Z_2$, and get
\begin{equation}
H_{H_2}''\sim -\frac{1}{\sqrt{2}}I_2+\frac{1+\sqrt{2}}{\sqrt{2}}X_2'+\frac{1-\sqrt{2}}{\sqrt{2}}Z_2'.
\end{equation}
Choosing $-X_2'$ as the second stabilizer and discarding $Z_2'$, we obtain an energy $-1-\sqrt{2}\approx -2.414$, which is significantly lower than the naive stabilizer value $-2$. Still, this is higher than the lowest eigenvalue of $H_{H_2}''$, which is $-2\sqrt{2}\approx -2.828$.

\subsection{Continuous gauging}

All the previous steps could be generalized by introducing a gauging parameter, which could vary continuously. This could be easily done with a general rotation $R_y(\theta)$ around the $\hat y$ axis.
The gauge transformation would then be given by:
\begin{eqnarray}
X'&=&R^\dagger_y(\theta)~X~R_y(\theta), \\
Z'&=&R^\dagger_y(\theta)~Z~R_y(\theta),
\end{eqnarray}
When $\theta=\pi/4$, we reproduce the previous special transformation as expected.

%When $\theta=0$, we have $X'=X$ and $Z'=-Z$, essentially the trivial transformation.
%It is not difficult to see that, $W(0)=W$. This is why we include a minus sign in the second line.
%An intriguing property of such a gauge transformation is that, the trajectory $(X(\theta),Z(\theta))$ never crosses the original point $(X,Z)$.
%To devise a gauge transformation connected to identity, perhaps we need to take the exponential in some proper way, or include the identity matrix %in the definition of $W(\theta)$.

Now we could repeat the whole calculation for the hydrogen molecule with the general transformation. By properly choosing the gauge parameters, we could further lower the energy from $-1-\sqrt{2}\approx -2.414$ to $-2.5$, even closer to the exact value $-2\sqrt{2}\approx -2.828$.

\section{Summary}

In this short note we employ the quantum strategy in the CHSH game to improve the stabilizer approximation for groundstates. The resulting approach turns out to be a gauging procedure for the Pauli operators $\{Z,X\}$. As expected, it significantly improves the performance when the original approximation deteriorates. The new stabilizer states in the gauged basis could then provide better initialization for further quantum algorithms~\cite{CAFQA,Cheng-2022}.

Such an approach also inspires a natural question: could it be enhanced to a universal computation model in the Heisenberg picture~\cite{Gottesman-1998}? Perhaps the answer is already hidden in the framework of the so-called ZX-calculus~\cite{ZX}, I guess.

%\section*{Note added:}

\section*{Acknowledgments}
I got the idea of gauging stabilizers when I was reading the series of blogposts ``{\it A brief introduction to quantum PCP conjecture}'', written in Chinese by the blogger ``{\bf Climber.pI}'' on his homepape `` \href{https://climberpi.github.io/}{\it Complexity Meets Quantum} ''. I don't know the author personally, and I am not even aware of his name. So I would like to take this opportunity to convey my appreciation to him.

%\newpage

\end{document}